# Proposal of Readout Electronics for CSNS-WNS BaF$_2$ Detector[*]

De-Liang Zhang (张德良)[1,2]　　Ping Cao (曹平)[1,3;1)]　　Qi Wang (王奇)[1,2]　　Bing He (何兵)[1,3]
Ya-Xi Zhang (张雅希)[1,2]　　Xin-Cheng Qi (齐心成)[1,3]　　Tao Yu (余滔)[1,2]　　Qi An (安琪)[1,2]

[1] State Key Laboratory of Particle Detection and Electronics, University of Science and Technology of China, Hefei 230026, China
[2] Department of Modern Physics, University of Science and Technology of China, Hefei 230026, China
[3] School of Nuclear Science and Technology, University of Science and Technology of China, Hefei 230027, China

**Abstract:** BaF$_2$ (Barium fluoride) detector is one of the experiment facilities at the under construction CSNS-WNS (White Neutron Source at China Spallation Neutron Source). It is designed for precisely measuring (n, γ) cross section with total 92 crystal elements and completely 4π steradian coverage. In this proposal for readout electronics, waveform digitizing technique with 1GSps sampling rate and 12-bit resolution is adopted to precisely capture the detector signal. To solve the problem of massive data readout and processing, the readout electronics system is designed into a distributed architecture with 4 PXIe crates. The digitized detector's signal is concentrated to PXIe crate controller through PCIe bus on backplane and transmitted to data acquisition system over Gigabit Ethernet in parallel. Besides, clock and trigger can be fanned out synchronously to each electronic channel over a high-precision distributing network. Test results showed that the prototype of the readout electronics system achieved good performance and cooperated well.

## 1 Introduction

Neutron radiation capture cross section, namely (n, γ) cross section, is used to describe the probability of radiation capture reaction between the incident neutron and the target nucleus. (n, γ) cross section is an important nuclear parameter, which plays a key role in the study of nuclear astrophysics, nuclide origin theory, nuclear energy development and nuclear waste disposal [1].

A 4π steradian detector composed of BaF$_2$ crystals is regarded as the best way to measure (n, γ) cross section. Several 4π steradian BaF$_2$ detectors have been constructed in the world, such as the barium fluoride detector (BFD) at Karlsruhe in Germany, the detector for advanced neutron capture experiment (DANCE) at Los Alamos National Laboratory in America, the total absorption calorimeter (TAC) at CERN and the gamma-ray total absorption facility (GTAF) at China Atomic Energy Science Research Institute. With these facilities, scientists obtained the (n, γ) cross section of many nuclides and made significant contributions to the nuclear research and application [2~5].

In China, an advanced spallation neutron source called China Spallation Neutron Source (CSNS) is under construction in Dongguan. It could produce wide band pulsed neutron beam, which is perfect for nuclear data measurement [6]. So a new white neutron source (WNS) pipe is being constructed in the opposite direction of the proton beam at CSNS and a new 4π steradian BaF$_2$ detector, namely CSNS-WNS BaF$_2$ detector, is planned to build to measure (n, γ) cross section. CSNS-WNS BaF$_2$ detector is composed of neutron pipeline, sample room, neutron absorber, BaF$_2$ crystals and photomultiplier tubes (PMTs) from inside to outside. It contains 92 BaF$_2$ crystals, each of which has a size of internal diameter of 20cm and thickness of 15cm, as shown in Fig. 1. Each crystal is coupled with an ultraviolet sensitive PMT for photoelectric conversion. BaF$_2$ detector's signal has a wide energy range from tens of keV to 10 MeV and wide frequency distribution from hundreds of Hz to about 120MHz.

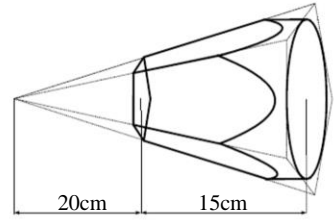

Fig. 1. BaF$_2$ crystal in CSNS-WNS BaF$_2$ detector

A readout electronics system is needed to capture the BaF$_2$ detector signal. To precisely measure (n, γ) cross section, the readout electronics should measure time of flight of the neutron, energy and multiplicity of the gamma rays. Moreover, the readout electronics should be able to discriminate the gamma rays from background (mainly alpha rays) by pulse shape discrimination methodology (PSD). To realize multi-parameter measurements at the same time, high-speed and high-resolution waveform digitizing technique is preferred to precisely acquire the signal's waveform, which contains complete physical information and could be used to extract any interested information. According to the physical requirements, a dynamic range about 500 in amplitude, energy resolution better than 1% @662keV and time resolution better than 1ns should be achieved for each electronics channel. To achieve these goals, a sampling rate up to 1GSps and an ENOB (Effective Number of Bits) up to 9bit below 100MHz for waveform digitizing should be realized. Therefore, an ADC (Analog to digital converter) chip with 1GSps sampling rate and 12-bit resolution is adopted to realize the waveform digitizing module.

[*] Supported by National Research and Development plan (2016YFA0401602) and NSAF (U1530111).
1) E-mail: cping@ustc.edu.cn

High-speed, high-resolution waveform digitizing leads to massive data readout and processing, 12Gbps raw data rate for each channel. To solve this problem, the readout electronics system is designed into a distributed architecture with 4 PXIe crates and 4 Gigabit Ethernets. Hardware-based trigger judgement and zero channel compression are introduced into FPGA (Field Programmable Gate Array) for decreasing the data rate. Moreover, high-precision synchronous clock and trigger networks should be designed to achieve synchronization for every electronics channel.

In this paper, a prototype of the $BaF_2$ detector readout electronics is proposed and developed. It contains analog circuit, waveform digitizers, synchronous clock and trigger network and data transmission network. To verify the performance of the readout electronics system, some tests were carried out.

## 2 Architecture of readout electronics

The architecture of the readout electronics is illustrated in Fig. 2. It mainly contains analog circuit, waveform digitizers, synchronous clock and trigger and data transmission network [7].

The whole system contains 92 channels. The readout electronics system is located about 20m away from the $BaF_2$ detector, so a 20m analog cable must be applied to transmit the analog signal from $BaF_2$ detector to the readout electronics for each channel. A high bandwidth, high slew rate, low noise preamplifier is used to receive and adjust the analog signal at output of the detector. At the backend, each analog output is divided into two signals, one is sent to field digitization module (FDM) for waveform digitizing, the other is sent to trigger module for trigger generating.

To achieve fast rising time and high SNR (signal-to-noise ratio), high bandwidth differential twisted-pair cable with shield should be chosen and the analog signal should be transmitted in differential mode. Correspondingly, the preamplifier is implemented with a fully differential amplifier to achieve single-ended to differential conversion. At the backend, a NIM based analog fan-out module (AFM) is designed to execute the signal deviation. The AFM is also implemented with a fully differential amplifier, which receives the differential signal from the cable and output a pair of differential voltage signals. The negative signal is sent to FDM and the positive signal is sent to trigger module. There are 8 AFMs in total and each AFM takes charge of 12 input channel. Because of the analog signal's wide frequency distribution, nearly from dc to 120MHz, dc coupling is introduced along the whole transmission route [8].

To digitize up to 92 detector channels, there are 46 FDMs, each of which contains two digitizing channels. The FDMs are designed as standard PXIe 3U plugins and settled in 4 PXIe crates. Multi-channel, high-speed, high-resolution waveform digitizing leads to huge data rate. To deal with the massive data readout and processing, high-bandwidth PCIe bus, distributed structure with 4 PXIe crates and 4 Gigabit Ethernets are introduced into the readout electronics system. Each crate contains a controller and several FDMs. To decrease the data rate, trigger judgement, zero channel compression and data buffering are performed with hardware on FDM. After acquired in FDM, the digitized crystal signal is concentrated to PXIe crate controller through the backplane PCIe bus and transmitted to data acquisition system over Gigabit Ethernet in parallel.

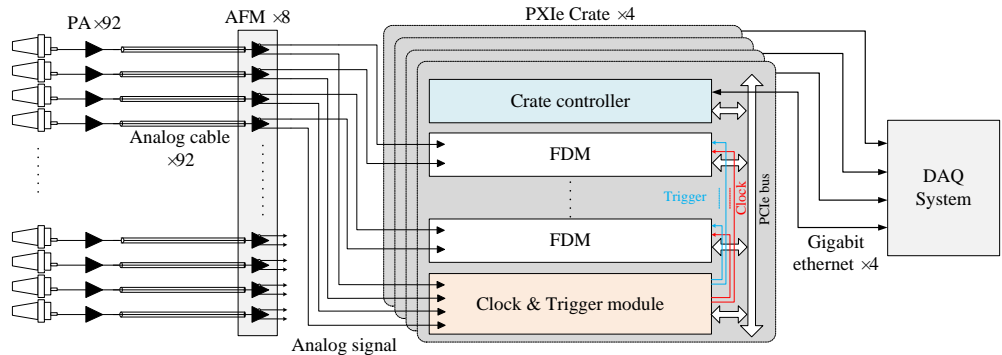

Fig. 2. Architecture of readout electronics system

High-precision synchronous clock network and trigger network should be designed to ensure all FDMs operating synchronously. To achieve high simplicity and precision, the clock network and trigger network are implemented deeply combining with the backplane star fan-out buses of the PXIe crate. As all the star fan-out buses are concentrated on the system timing slot, we have to integrate the clock network and trigger network as an inseparable network.

The clock and trigger network has a twin-stage structure. In the first stage, global trigger and clock module (GTCM) produces and distributes clocks and triggers to local trigger and clock modules (LTCM) among PXIe crates. In the second stage, LTCMs fan out triggers and clocks separately to other electronics plugins (mainly FDMs) within the same crate, acting as buffers.

## 3 Waveform digitizer

A waveform digitizer called field digitization module (FDM) is designed for acquiring the waveform of the detector's signal. It is the crucial part of the readout electronics. As illustrated in Fig. 3, the FDM is designed as a standard PXIe 3U plugin. It is mainly composed of an ADC, a FPGA, two DDR3 memories, analog conditioning circuit, clock circuit, trigger circuit, PCIe interface and USB interface. The USB interface is spare for debugging. [7].

To achieve 1GSps sampling rate and 9-bit ENOB, an ADC based on folding interpolation from TI (ADC12D1000) is adopted to digitize the input signal on FDM. It contains 2 channels and each channel is typically operating under 1GSps sampling rate with 12-bit resolution. It requires differential inputs and could provide 1.25V common-mode voltage for dc-coupling. It is powered by 1.9V single supply. With the ADC chip, an ENOB (Effective Number of Bits) higher than 9bit and power dissipation about 1.7W for each channel could be achieved [9].

To achieve dc-coupling and accommodate the ADC's input, a fully differential amplifier is used to buffer the analog input and achieve single-ended to differential conversion. A π-type resistors network is applied for terminating and attenuating the input signal. For further improving the SNR, an anti-aliasing filter is settled at the ADC's analog input.

The clock for FDM is buffered in from one of the PXIe's backplane differential star bus, PXIe_DstarA, which is drived in LVPECL signal and could achieve high performance with jitter lower than 3ps and skew lower than 150ps. A PLL (Phase Locked Loop) with jitter cleaner (LMK04821) [10] is used to multiply the clock's frequency and filter out the jitter noise, to achieve the high-precision 1GHz sampling clock. Besides, a local oscillator is adopted on FDM for PLL configuration after power-up and an external clock interface is spare for debugging.

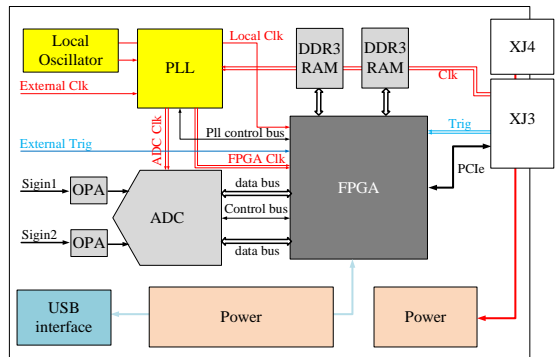

Fig. 3. Schematic of the field digitization module

A high speed Xilinx Kintex-7 series FPGA (XC7K160T-3FFG676) is adopted as the controller on FDM. The FPGA takes charge of data receiving and processing in real time. For each channel, FPGA receives the 12Gbps data from ADC, deserializes it into 96 bit with 125MHz and caches in FIFO (First-In First-Out) memory. Once a valid trigger signal occurs, the effective data segment is selected and buffered into next FIFO. After trigger judgement, zero channel compression algorithm is implemented for further decreasing the data rate. The trigger is input from PXIe_DstarB, another differential star bus on the backplane of the PXIe crate, which is drived in LVDS signal and could achieve high performance with jitter lower than 3ps and skew lower than 150ps. Finally, the data is packaged with trigger ID, buffered in DDR3 memories and sent to PXIe crate controller by PCIe DMA. Two DDR3 memories operate in ping-pang mode, each of which has a capacity up to 4Gbit.

## 4 Clock and trigger network

A synchronous clock and trigger network is designed to ensure the whole readout electronics system operating synchronously. It contains the generation and distribution of the clock and trigger signals. By analysis and simulation, to achieve an ENOB up to 9bit and time resolution better than 1ns, the 1GHz sampling clock should achieve high performance with jitter lower than 660fs and skew far less than 1ns. It requires the system clock from the clock and trigger network with jitter lower than 100ps and skew far less than 1ns.

### 4.1 Clock and trigger generation

The clock signal is 100MHz, which is generated with a local oscillator on GTCM (Global Trigger and Clock Module). A PLL (LMK04821) is used to filter out the jitter noise and fan out the clock signal to internal logic and the GTX module of the FPGA. After generated, the clock is distributed with high speed serial communication and clock data recovery (CDR) technique with the GTX module.

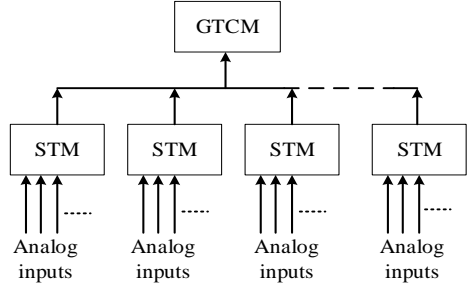

Fig. 4. Trigger generation network

The trigger signal is generated by two steps, as shown in Fig. 4. For the first step, the STM (Sub Trigger Module) receives the analog signals and generates sub trigger signals, which contain the analog energy sum and the over-threshold channels information. The analog energy sum is achieved be analog integration and analog summation by operational amplifiers. It is sent to GTCM through coaxial cable through the front panel. The over-threshold channels information is achieved by multi-channel discriminators and sent to GTCM through PXIe_DstarC, a differential star bus from every peripheral slot to the system timing slot on the PXIe backplane. For the second step, the GTCM receives the sub trigger signals and generates the global trigger, according with the time window of the beam, multiplicity and energy threshold. There are 6 STMs in total and each STM takes charge of 16 analog inputs. The STMs are settled in the same PXIe crate with the GTCM.

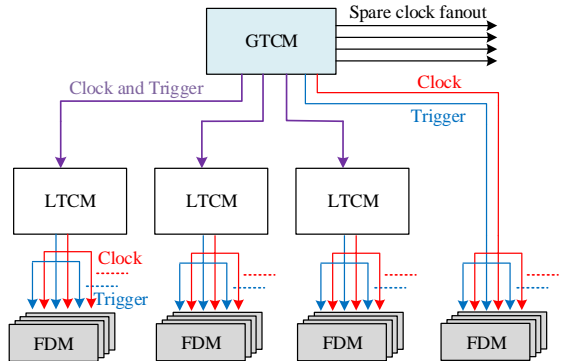

Fig. 5. Clock and trigger distribution network

Both the clock and trigger are generated on GTCM. After generated, the clock and trigger are distributed to the whole readout electronics system through a common distribution network. The clock and trigger distribution network has a tree-like topology with a twin-stage structure, as shown in Fig.5. GTCM is responsible for generating global trigger and clock, then distributing them to LTCMs. LTCM is responsible for fanning out clocks and triggers within the same crate. GTCM is located at the system timing slot of the master PXIe crate while three LTCMs are located at the system timing slots of the slave PXIe crates.

### 4.2 Clock and trigger distributing between crates

To achieve high simplicity and high precision, the clock and trigger distribution networks are implemented deeply combining with the backplane star fan-out buses of the PXIe crate. As all the star fan-out buses are concentrated on the system timing slot, the global clock and trigger distribution have to share one system timing slot plugin, GTCM. It results in too many channels integrated on GTCM and too many cables connected to the front panel of GTCM, which is disobedient to the limited area of the front panel of GTCM. To deal with this conflict, high speed serial communication and clock data recovery (CDR) technique are adopted to transmit the clock and trigger together.

Furthermore, single-ended transmission with coaxial cable is used to replace traditional differential transmission for further simplifying the distribution network and micro-miniature coaxial (MMCX) connectors are used to save panel space. Pre-emphasis and equalization technique are used to ensure the performance of the single-ended signal through a long coaxial cable. So, one coaxial cable could realize the clock and trigger distribution between crates at the same time, which makes the readout electronics system more concise. Modulation and demodulation of the clock and trigger are achieved by GTX module in FPGA. The baud rate on the coaxial cable is 1GHz and the coaxial cable is about two meters long. The clock and trigger are recovered on LTCM. A high performance recovered clock is obtained through a PLL with jitter cleaner. The distributed clock is 100MHz in frequency [7].

### 4.3 Clock and trigger fanning out inside a crate

High performance backplane star fan-out buses of the PXIe crate are used to fan out clocks and triggers inside a crate. PXIe crate provides three sets of backplane star buses. The PXIe_DstarA is a differential star fan-out bus drived by LVPECL signal from the system timing slot to all peripheral slots. The PXIe_DstarB is a differential star fan-out bus drived by LVDS signal from the system timing slot to all peripheral slots. The PXIe_DstarC is an oppositely directed star bus from every peripheral slot to the system timing slot. They all could achieve high performance with jitter lower than 3ps and skew lower than 150ps.

So the PXIe_DstarA is used to fan out clocks, the PXIe_DstarB is used to fan out triggers and the PXIe_DstarC is used to upload the sub-trigger information. With the clock and trigger distribution network, a high performance, highly concise readout electronics system is achieved [12].

Beside, a spare clock fan-out network is contained in the clock and trigger network for system expansion. It is described in detail in reference [11].

# 5 Tests and Verification

Several tests were carried out to verify the key performance of the readout electronics, including the waveform digitizing test, the clock performance test, data transmission test and cosmic ray test.

### 5.1 Waveform digitizing performance

The waveform digitizing performance test was carried out with the FDM, which included static performance test and dynamic performance test.

Sine waveforms with the frequency of 2.4MHz were used for the static characteristics evaluation of FDM. The static parameters were calculated by the code density method and ten million sampled points were used. Test results showed that, the INL (Integral Non-Linearity) of the FDM was range from -2.5 LSB (Least Significant Bit) to +1.5 LSB and the DNL (Differential Non-Linearity) was range from -0.20 LSB to +0.20 LSB. The offset error was -0.99 LSB and the gain error was -0.02 LSB. Perfect static performance was achieved, which is similar to the parameters on the ADC's datasheet.

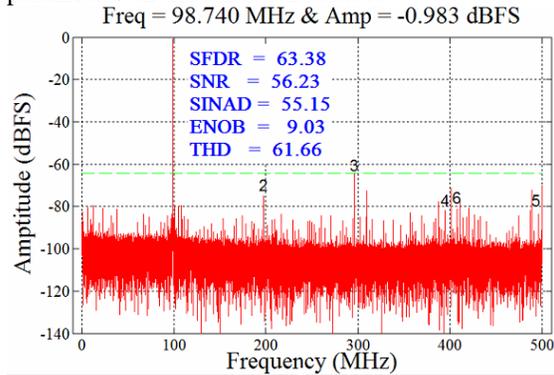

Fig. 6. Results of dynamic performance tests

Sine waveforms with frequencies from 2.4 MHz to 198 MHz were used for the dynamic performance tests. Analysis was performed via FFT (Fast Fourier Transform) and 65536 sampled points were involved. Test results showed that the FDM achieved an ENOB above 9bit with the frequency below 100MHz, which could satisfy the waveform digitizing requirement for the $BaF_2$ detector signal. Dynamic parameters at 98MHz are illustrated in Fig. 6.

### 5.2 Data transmission capability

Block diagram of the data transmission network is shown as Fig. 7, the transmission route covered from the FDM to DAQ computer. PCIe DMA was adopted for data transmitting between FDM and the crate controller, while Gigabit Ethernet was adopted from the crate controller to the DAQ computer. Data transmission test was performed under linux operating system.

Test result showed that an instantaneous data rate of 942Mbit/s was achieved, which is more than 90% of the theoretical value. Four Gigabit Ethernets could well meet the data transmission requirement.

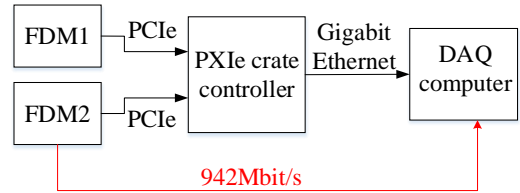

Fig. 7. Block diagram and result of data transmission test

### 5.3 Clock distribution performance

Clock performance test was performed to evaluate the clock performance after the clock and trigger network. The test block diagram is shown in Fig. 8. GTCM and the test points on it were used to simulate the clock and trigger distribution network. The loopback coaxial cable simulates the clock and trigger distribution process between crates. TP5 represents the terminal point of the clock distribution network.

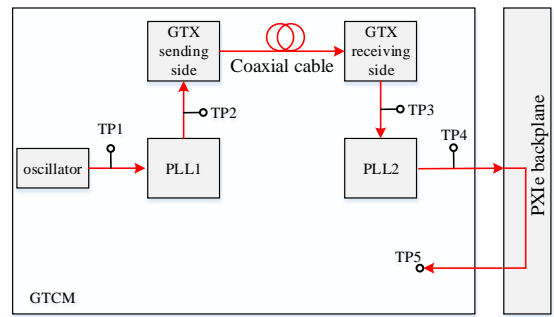

Fig. 8. Block diagram of clock distribution test

The clock performance was obtained by a Tektronix DPO5104 oscilloscope. The jitter of the distributed clock was characterized in TIE (Time

Interval Error) at a population of $10^5$, which is enough to get a highly accurate statistical result.

Test results are shown in Table 1. A low jitter about 12ps was achieved at the terminal point of the clock distribution network.

Table 1 Results of the clock distribution test

| Test point | TIE |
|---|---|
| TP1 | 17.2ps |
| TP2 | 8.3ps |
| TP3 | 0m coaxial cable: 51.8ps |
| | 1m coaxial cable: 49.7ps |
| | 2m coaxial cable: 56.4ps |
| | 3m coaxial cable: 61.2ps |
| TP4 | 10.3ps |
| TP5 | 12.2ps |

Skew of the clock distribution network was also obtained. As Fig. 9 shows, a low skew about 200ps was achieved between different slots within a PXIe crate.

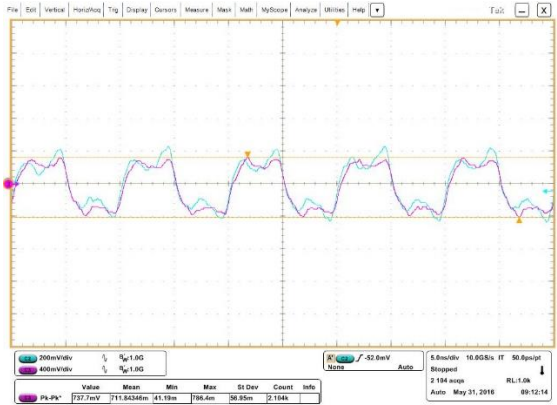

Fig. 9. Skew clock distribution network between different slots within a PXIe crate

The clock performance test results showed that a high performance with jitter about 12ps and skew about 200ps was achieved for the system clock. It could well meet the requirements for ENOB and time resolution.

### 5.4 Cosmic ray experiment

Cosmic ray experiment was carried out to evaluate the cooperation of the readout electronics system. The experiment platform contained $BaF_2$ crystal, PMT and high-voltage module, preamplifier, AFM, FDM, PXIe crate and the data acquisition software, as illustrated in Fig. 10.

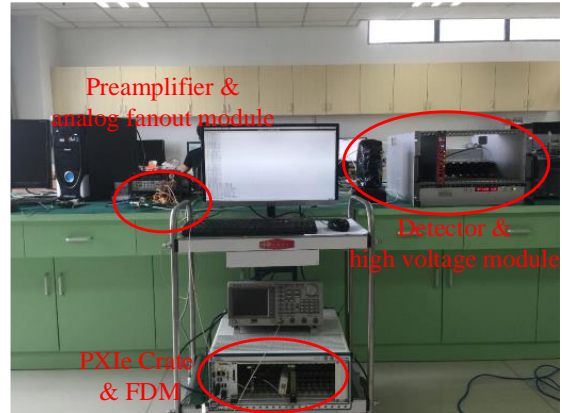

Fig. 10. The cosmic ray experiment platform

Typical waveforms of the cosmic ray and the alpha ray were acquired by the readout electronics system, as shown in Fig. 11 and Fig. 12, which indicates that the readout electronics system cooperate well.

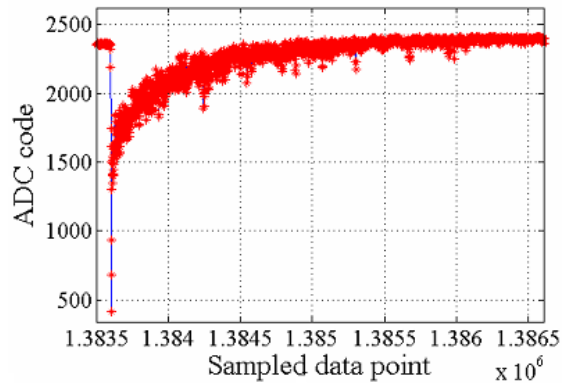

Fig. 11. Typical waveform of the cosmic ray

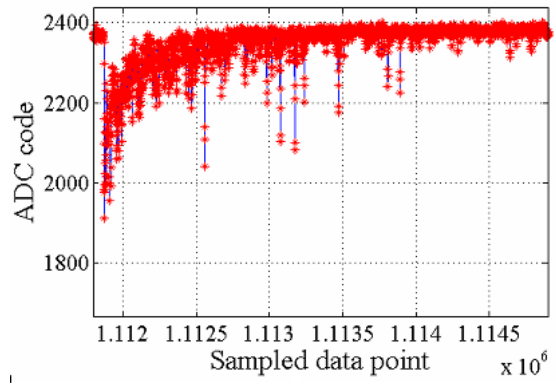

Fig. 12. Typical waveform of the alpha ray

The cosmic ray's waveform was precisely captured by the $BaF_2$ detector readout electronics. Obviously, the cosmic ray's waveform had fast component but the alpha ray had not. Pulse shape discrimination methodology could be used to discriminate the γ-rays from the alpha rays. Four sampling points were captured at the leading edge, which was similar to that captured by oscilloscope and could be used to obtain the time of flight of the neutron. Finally, the energy of the γ-ray could be figured out by digital waveform integration.

## 6 Conclusion

A prototype of the $BaF_2$ detector readout electronics was proposed and developed. It has a distributed structure with 4 PXIe crates and 4 Gigabit Ethernets. To precisely measure (n, γ) cross section, high speed, large dynamic range waveform digitizing technique with 1GSps sampling rate and 12-bit resolution is used to precisely acquire the signal's waveform. High-precision synchronous clock and trigger network is design to ensure all FDMs operating synchronously.

Test results showed that the FDM achieved perfect static performance and dynamic performance. The data transmission network achieved a transmission rate more than 90% of the theoretical value. The system clock achieved a high performance with jitter about 12ps and skew about 200ps. These performances could well meet the requirements of $BaF_2$ detector's signal readout. Cosmic ray experiment indicated that the readout electronics system cooperated well. The cosmic ray's waveform was precisely captured by the $BaF_2$ detector readout electronics.

**Acknowledge:**
*The authors gratefully acknowledge all members of CSNS-WNS for their earnest support and help during the work.*